\documentclass[aps,preprint]{revtex4}
\usepackage{latexsym}
\usepackage{amsfonts}
\usepackage{amsmath}
\usepackage{bm}
\usepackage{amssymb}
\usepackage{epsfig}

\begin{document}
\title{Analyzing intramolecular dynamics by Fast Lyapunov Indicators}

\author{E. Shchekinova$^1$, C. Chandre$^2$, Y. Lan$^1$ and T. Uzer$^1$}

\affiliation{$^1$ Center for Nonlinear Science, School of Physics,
Georgia Institute of Technology, Atlanta, Georgia 30332-0430, U.S.A.\\
$^2$ Centre de
Physique Th\'eorique - CNRS, Luminy - Case 907, 13288 Marseille
cedex 09, France}

\begin{abstract}
We report an analysis of intramolecular dynamics of the highly
excited planar carbonyl sulfide (OCS) below and at the
dissociation threshold by the Fast Lyapunov Indicator (FLI)
method. By mapping out the variety of dynamical regimes in the
phase space of this  molecule, we obtain the degree of regularity
of the system versus its energy. We combine this stability
analysis with a periodic orbit search, which yields a family of
elliptic periodic orbits in the regular part of phase space an a
family of in-phase collinear hyperbolic orbits associated with the
chaotic regime.
\end{abstract}

\maketitle

\section{Introduction}

Multidimensional, complex systems tend to overwhelm the researcher
with data: think of fluid dynamical data from the oceans or the
millions of trajectories that today's powerful computers can
generate in an instant. In many areas of science and engineering,
experimental techniques to observe real-time dynamical phenomena
have also developed at a pace far more rapid than the theory
required to make sense of such data. In chemistry, a recent
example is the rapid development of techniques to study single
molecules, such as proteins~\cite{lu98,onuc99}.

For low-dimensional systems, the geometric framework of dynamical
systems theory has provided a way of understanding these data.
This point of view, first espoused by Poincar\'e, asks about the relationship
between all possible trajectories, rather than
the evolution of individual trajectories. This
leads immediately to the notion of {\em phase space structure} as a key
notion for making sense of the many and varied regimes that a
nonlinear dynamical system can exhibit. However, techniques for
mining high-dimensional plethora of data for any underlying
structures and geometry have thus far lagged badly behind our
ability to generate it.

For Hamiltonian systems with two degrees of freedom, a Poincar\'e
section constructed by a plane section (of dimension 2) of a set
of individual trajectories lying on the energy surface of
dimension 3 is able to give a clear picture of the dynamics, e.g.,
of regular versus chaotic regions of phase space. Analyzing
the dynamics of higher- dimensional systems by extrapolating
this technique is not straightforward since the Poincar\'e section
has dimension 4 for systems with three degrees of freedom. One
method to understand  high-dimensional systems has been developed by
Laskar~\cite{lask90,lask93,lask99} and relies on visualizing the dynamics in
the frequency space. For instance, for three degrees of freedom,
this frequency space is of dimension 2 which makes its analysis
tractable. It is known as Frequency Map Analysis (FMA) because it is  based on
 extracting the principal  frequencies of quasiperiodic
trajectories. Clearly, this method is well-suited for nearly
integrable systems whose phase space contains many invariant
tori. This method has also been extended to weakly chaotic regimes by
computing  a diffusion coefficient in frequency. The FMA has been
first developed in celestial mechanics, and has subsequently been applied in
various other fields, like atomic physics~\cite{vonM96,vonM97}, particle
accelerators~\cite{duma93,lask96,robi00} and
chemistry~\cite{losa98,milc98}. For
strongly chaotic systems, the notion of frequency and diffusion in
frequency are not well defined, so any Fourier-based analysis like
this one must
be used with extra caution.

In order to understand the dynamics of high-dimensional systems in
the strongly chaotic regime, the computation of Lyapunov exponents
based on the linearized flow has been used
extensively~\cite{Blich83}. However, Lyapunov exponents being
infinite-time quantities, the long computational times required
for their reliable computation are not suited for an extensive
analysis of phase space. Geometrical approaches based on the
geodesic deviation equations to study dynamics of many-dimensional
continuous dynamical systems have been developed in~\cite{cipr98}.
Also noteworthy are the investigations of Berry and coworkers
concerning the non-uniformity of the dynamical properties of
Hamiltonian systems representing atomic clusters with up to 13
atoms. In particular, they explored how regular and chaotic
behavior may vary locally with the topography of the potential
energy surfaces (PES)
\cite{Berry88a,Berry91,berry92,Berry92a,Berry93a,Berry93b,Berry93c,Berry96a,Berry98,Komatsuzaki99,Komatsuzaki02a}.
By analyzing local Lyapunov functions and Kolmogorov entropies,
they showed that when systems have just enough energy to go
through a saddle in the potential energy surface, the system's
trajectories become collimated and regularized through the saddle
regions, developing approximate local invariants of the motion
different from those in the potential well. Recently, numerical
methods based on linear stability analysis have been designed to
analyze trajectories of Hamiltonian systems, which were obtained
by numerical integration of the flow. The aim of these methods is
to obtain pictures of the phase space and highlight relevant
structures~\cite{guzz02,megno1,megno2}. It has been shown that the
relevant information on a typical trajectory is obtained by
integrating the flow for a short time. For instance,
distinguishing between regular and chaotic trajectories, or even
between weakly and strongly chaotic trajectories is possible after
a short integration of the equations of motion. The Fast Lyapunov
Indicator (FLI) method is one of these highly practical methods.

In this article, we use the FLI method to analyze the internal
dynamics of the carbonyl sulfide (OCS) molecule. In its planar and
rotationless configuration, this molecule has three coupled modes
(two stretching modes and one bending mode). We show how this
method helps to visualize the dynamics of highly excited molecules
without resorting to dimensional reduction schemes with their
attendant flaws and loss of information. In particular, the Fast
Lyapunov Indicator method shows global pictures of the dynamics,
highlighting regular and chaotic zones. We combine the analysis of
the dynamics of OCS  with a determination of the main periodic
orbits. Our numerical investigations yield  the following
picture~: The regular region is located in the center of the
configuration space and is characterized by a family of elliptic
periodic orbits. The strongly chaotic region is entered when the
CS stretching mode is highly excited and this region is
characterized by a set of collinear hyperbolic periodic orbits
which have an additional property of being 'in-phase' (meaning
that the bending mode is frozen and the two stretching modes are
vibrating with the same frequency).

In Sec.~\ref{sec1}, we briefly give the explicit expression of the flow of the OCS. In Sec.~\ref{sec2}, we explain
 the basics of the Lyapunov indicator method. The numerical results are given in Sec.~\ref{sec3}.

\section{The model}
\label{sec1}
The classical model of the planar (rotationless) carbonyl sulfide OCS molecule
has been studied in details in
Refs.~\cite{cart82,davi84,dawa84,davi85,mart87,Tmart87}. The
coordinates of this system are two interatomic distances $R_1=d(C,S)$,
$R_2=d(C,O)$, the bending angle of the molecule
$\alpha=\widehat{OCS}$, and three momenta $P_1$, $P_2$, $P_\alpha$
which are conjugate to $R_1$, $R_2$ and $\alpha$, respectively . We note that the third interatomic distance $R_3=d(O,S)$ is expressible as a function of $R_1$, $R_2$ and $\alpha$~:
$$
R_3=\left(R_1^2+R_2^2-2R_1 R_2\cos\alpha\right)^{1/2}.
$$
The Hamiltonian for this system is
\begin{equation}
H(R_1,R_2,\alpha,P_1,P_2,P_\alpha)=T(R_1,R_2,\alpha,P_1,P_2,P_\alpha)+V(R_1,R_2,\alpha),
\end{equation}
where $T$ is the kinetic part of the Hamiltonian and $V$ is the potential. The kinetic part
has the form~\cite{bunk62}
\begin{eqnarray*}
T&=&\frac{\mu_1}{2}P_1^2+\frac{\mu_2}{2}P_2^2+\mu_3P_1P_2\cos\alpha\\
&& +P_\alpha^2\left( \frac{\mu_1}{2R_1^2}+\frac{\mu_2}{2R_2^2}-\frac{\mu_3\cos\alpha }{R_1 R_2}\right)-\mu_3P_\alpha \sin \alpha \left(\frac{P_1}{R_2}+\frac{P_2}{R_1}\right),
\end{eqnarray*}
where $\mu_i$ are the reduced masses.
The analytic expression
of the potential has been proposed based on existing
experimental data~\cite{foor75} and can be summarized as :
\begin{equation}
V(R_1,R_2,\alpha)=\sum\limits_{i=1}^3 V_i(R_i)+V_I(R_1,R_2,R_3),
\end{equation}
where $V_i$ are Morse potentials for each diatomic pair
$$
V_i(R)=D_i\left(1-e^{-\beta_i(R-R_i^*)}\right)^2,
$$
and $V_I$
is the interaction potential of the Sorbie-Murrell form given in
Ref.~\cite{cart82}~:
$$
V_I=P(R_1,R_2,R_3)\prod_{i=1}^3 \left(1-\tanh \gamma_i(R_i-R_i^{(0)})\right),
$$
where $R_i^{(0)}$ are the equilibrium distances of the planar OCS which is a collinear ($\alpha=\pi$) configuration~: $R_1^{(0)}=2.9508$ (in atomic units) $R_2^{(0)}=2.2030$ and $R_3^{(0)}=R_1^{(0)}+R_2^{(0)}$. The function $P$ is a quartic polynomial in each of its variables. All the coefficients of the potential are provided in Ref.~\cite{cart82}.

The planar OCS is an example of a general triatomic molecule
with three strongly coupled modes~: OC and CS stretching modes, represented by $R_1(t)$ and $R_2(t)$ and a
bending mode, represented by $\alpha(t)$. Any perturbation
introduced into one of the modes will be redistributed among the two
other modes. It was shown in Ref.~\cite{cart82} that the dynamics of the system
is highly irregular at the energies close to the dissociation
threshold, which occurs at $E=0.1$ (in atomic units).

\section{The Fast Lyapunov Indicator method}
\label{sec2}
The  Fast
Lyapunov Indicator (FLI) method was introduced in Ref.~\cite{froe97} and
rigorous results can be found for the case of near integrable systems in Ref.~\cite{guzz02}. The
method is similar to the computation of finite time Lyapunov
exponents \cite{okushima}.
Given a $d$-dimensional flow
$$
\frac{d {\bf x}}{dt} = {\bf f}({\bf x}),
$$
we are looking at the evolution of a vector ${\bf v}$ which is given by the tangent flow
$$
\frac{d{\bf v}}{dt}=Df({\bf x}) \cdot {\bf v},
$$
where $Df$ is the matrix of the variations of the flow given by the
velocity field ${\bf f}$, i.e.\ $[Df]_{ij}=\partial f_i /\partial
x_j$. We integrate the above system of equations starting with initial conditions ${\bf x}_0$ and ${\bf v}_0$. In principle, one should consider the dynamics of the $d\times d$ Jacobian matrix $J(t)$ which is given by
$$
\frac{d J}{dt}=Df({\bf x}) J,
$$
(where $J(0)$ is the $d\times d$ identity matrix), in order to study the stability of a given trajectory~\cite{Bchaos}.
The evolution of ${\bf v}$ is thus given by ${\bf v}(t)=J(t){\bf v}_0$. However, for practical purposes, we only
integrate the equations for ${\bf v}(t)$ starting with a given ${\bf v}_0$ fixed once and for all. We will omit the
 dependence on ${\bf v}_0$ which will be a valid assumption for large time since most of the vectors ${\bf v}_0$ will
  follow the dynamics of the most unstable eigenvector of $J(t)$.  \\
The Lyapunov indicator is based on the computation of the dynamical variable
$\phi(t)$ which is defined as follows:
\begin{equation}
\phi(t;{\bf x}_0)=\max_{0\leq t'\leq t}\log \Vert {\bf v}(t';{\bf x}_0)\Vert ,
\end{equation}
where ${\bf v}(t;{\bf x}_0)$ is a tangent vector of the flow
at time $t$ for the trajectory starting with initial conditions ${\bf x}_0$.
The Lyapunov indicator $\phi(t;{\bf x}_0)$ is a monotonically increasing function of time. Not only can it distinguish regular from chaotic motion but is
also an indicator of resonant and nonresonant trajectories
for nearly integrable systems~\cite{guzz02}.

For instance, a chaotic trajectory is characterized by an
exponential separation of nearby trajectories. The resulting
Lyapunov indicator grows linearly in time (the coefficient of this
growth will be asymptotically the largest Lyapunov exponent). In
the regular region, a good model is provided by considering an
integrable Hamiltonian with action-angle coordinates $({\bf
A},{\bm \varphi})\in \mathbb{R}^d\times \mathbb{T}^d$ where
$\mathbb{T}^d$ is the $d$-dimensional torus. The Hamiltonian
writes $H_0=H_0({\bf A})$ and the equations of motion are the
following ones
\begin{eqnarray*}
&& \frac{d{\bf A}}{dt}={\bf 0},\\
&& \frac{d{\bm \varphi}}{dt}=\frac{\partial H_0}{\partial {\bf A}}={\bm \omega}({\bf A}).
\end{eqnarray*}
The tangent flow is given by
$$
\frac{d{\bf v}}{dt}=\left(
\begin{array}{cc}
        0 & 0 \\
        {\bm \omega}'({\bf A}) & 0
\end{array} \right) \cdot {\bf v},
$$
where ${\bm \omega}'({\bf A})$ is a $d\times d$ matrix whose elements are $\partial \omega_i /\partial A_j$.
Since ${\bf A}$ is constant along a trajectory, the evolution of the vector ${\bf v}$ is linear in time. Therefore,
 the Lyapunov indicator evolves like
$$
\phi(t)\approx \log t.
$$
More precise results are obtained in the nearly integrable
regime~\cite{guzz02}. There is a linear growth in the chaotic region
and a logarithmic growth in the regular region. For $t$ large enough,
the Fast Lyapunov indicator makes a clear distinction between regular
and chaotic trajectories as we will show  numerically for the OCS
molecule. Our numerical observation is that this indicator achieves
this distinction very early in time compared with other existing methods.

The maps of the dynamics are obtained by examining the values of this Lyapunov indicator at a fixed time as
 a function of initial conditions.
$$
{\bf x}_0 \mapsto \phi(t;{\bf x}_0).
$$
In spirit, this method is very similar to Frequency Map Analysis~\cite{lask92,lask99} where instead of a Lyapunov
indicator, a diffusion in frequency is plotted as function of the initial conditions. The main feature of the FLI
 method is that it reveals the important phase space structures which makes the method an appropriate tool for the
 investigation of the classical phase space of highly excited molecules.

\section{Numerical Results}
\label{sec3} We investigated the phase space structures including
existing resonances and periodic trajectories and analyzed the
stability properties for different regions in the
configuration space of OCS. \\
In what follows we consider trajectories with initial conditions
in the configuration space $(R_1,R_2)$, i.e.\ we consider
initially that $P_1=P_2=P_\alpha=0$ and the initial value of the
angle $\alpha$ is determined by the energy integral
$V(R_1,R_2,\alpha)=E$. The  FLI as a function of time is computed
in the time interval $t\in [0, 1]$ (the unit of time is one
picosecond), which is long enough for the system to show the
characteristic dynamics. The integration of the equations of
motion is carried out using a standard variable order
Adams-Bashforth-Moulton PECE solver. Figure~\ref{fig:FLIt1}
depicts the variations of the Lyapunov indicator $\phi$ as a
function of time for four distinct trajectories at an energy
$E=0.09$ (in atomic units, where the value of the energy is given
with respect to the equilibrium energy)~: one periodic, one
quasiperiodic, one weakly chaotic and one strongly chaotic orbit.
A projection on the plane $(R_1,R_2)$ of each of these four
trajectories computed for $t\in [0, 1 ]$ is depicted on
Fig.~\ref{fig:FLIt1cs}. These figures show the power of the Fast
Lyapunov indicator: Not only can it distinguish between regular
and chaotic trajectories but also between resonant and
non-resonant regular trajectories and between weakly and strongly
chaotic trajectories. To set the scale, this distinction is made
as early as at time $t=1$. As a comparison, the molecule has a
period of 0.063 in the periodic case (the periodic orbit $O_a$,
see below), i.e.\ $t=1$ corresponds to about 15 periods of
oscillation of the molecule.

Fields of Fast Lyapunov Indicators were computed on an
equally spaced grid of initial conditions in the configuration
space $(R_1,R_2)$ for different energies. The computations for the FLI are carried
out on the time interval $t\in [0, 1]$.  For each initial condition
$(R_1,R_2)$, the maximum value attained by FLI during the
time interval $[0,1]$ ps is plotted. The result is  the Lyapunov field for a given value of the energy.
Figure~\ref{fig:FLIf1} depicts the Lyapunov field for $E=0.09$. The dark regions of this Lyapunov field
 are associated with regular regions and the white regions with chaotic trajectories.\\

In order to check these results, we  computed the main
periodic orbits and their stability at this energy.
We found these  periodic orbits by a novel variational method
which provides a very robust determination of periodic orbits of
flows~\cite{lan1,lan2}. This method can  determine periodic orbits of
flows regardless of their stability properties (elliptic or hyperbolic
periodic orbits). A brief description follows. \\
We start from a point  in the configuration subspace and evolve it for
some time. We take $R_1=3$ to be our Poincar\'{e} section by noticing that
most of the time the orbit intersects this hyperplane. The flow becomes a map
on the Poincar\'{e} section. As the symbolic dynamics of this system is yet unexplored, to obtain
cycles of different topological length, we
look for the near-recurrence of the map on the section for one iteration,
two iterations, three iterations, etc. The resulting orbit segment
is represented by a
discrete set of points and its frequency components are obtained by a
Fast Fourier Transform. After removing the high-frequency part, we
transform the data back to the phase space and obtain a closed smooth
loop, which becomes  our starting guess for a periodic orbit. A Newton descent flow will
drive the orbit toward a genuine periodic orbit by penalizing the discrepancy
between the approximate flow and the true flow along the loop~\cite{lan1,lan2}.
Table~I lists all the elliptic periodic orbits we have  obtained using this method~: Two projections of each periodic orbit are given, one on the plane $(R_1,P_1)$ and the other on the plane $(R_2,P_2)$. The period $T$ is also given in picosecond as well as the value of the resonance $m:n:k$. These integers are computed using the following procedure~: We compute by frequency analysis the main frequencies $\omega_1$, $\omega_2$ and $\omega_3$ of the three signals, respectively, $z_1(t)=R_1+iP_1$, $z_2(t)=R_2+iP_2$ and $z_3=\alpha+iP_\alpha$. The integers $m$, $n$, $k$ are such that $\omega_1/\omega_2=m/n$, $\omega_2/\omega_3=n/k$.
Each of these periodic orbits intersects the configuration space at two different points. These points are represented on Fig.~\ref{fig:FLIf1}.
Our main observation is that the center of the configuration space is weakly chaotic and characterized by two elliptic periodic orbits $O_a$ and $O_b$.\\
We have also computed the main hyperbolic periodic orbits of  OCS. The
intersection of these periodic orbits with the configuration space is
plotted on Fig.~\ref{fig:FLIf1}. We notice that the boundary of the
configuration space which mostly chaotic is characterized by
hyperbolic periodic orbits. Moreover, by frequency analysis we show
that these hyperbolic periodic orbits are collinear ones, i.e., the
bending mode is frozen and the stretching modes are in-phase ( $1:1:0$
resonances).




We note that the chaotic trajectories appear when $R_1$ is large or $R_1$ is small. Therefore the stretching mode $CS$ appears to lead the dynamics of the molecule. Moreover since in-phase collinear hyperbolic periodic orbits appear when $R_1$ is large, these type of periodic orbits are likely to lead the chaotic behavior near the dissociation threshold.

The resulting data on FLI values for
all the initial conditions corresponding to the different
dynamical regimes allow one to introduce a classification of the
orbits and to compute the percentage of the regular and irregular
orbits for the system for different values of the energy and to compare
those results with an existing classification. In Ref.~\cite{cart82}
the classification was done by using the behavior of the  Lyapunov exponents
for microcanonically chosen trajectories at different energies. Here, we examine the evolution of FLI curves for different
initial conditions in the configuration space.
As a result the threshold value of the FLI can be introduced for
filtering regular and irregular motions. We have noticed that for
the majority of trajectories studied the motion is chaotic or
weakly chaotic above the value $\phi_c=10$ (which is evaluated at $t=1$).

Figure~\ref{fig:FLIf2} depicts the Lyapunov fields for six
different values of the energy,  from $E=0.05$ up to the
dissociation energy $E=0.1$. One can observe the evolution of the
global stability of the system as the energy is increased. Dark
regions correspond to low values of FLI, that is to say, to
regular regions, whereas light regions are associated with high
values of the FLI where chaotic trajectories are predominant. The
color scale is chosen to be the same for each energy in order to
observe in a clearer way the changes in phase space as energy
increases. The FLI field for the energy $E=0.05$ consists of a
dark region in the center associated with a set of elliptic
periodic and quasiperiodic orbits with frequency ratios
$\omega_1/\omega_{\alpha}=2,~\omega_2/\omega_{\alpha}=2$ where
$\omega_1$, $\omega_2$ and $\omega_\alpha$ are   the main
frequencies of the signals $R_1(t)+iP_1(t)$, $R_2(t)+iP_2(t)$ and
$\alpha(t)+iP_\alpha (t)$, respectively. Smaller dark regions are
associated with higher order resonances. Most of trajectories are
quasiperiodic for this energy since the values of the FLI
$\phi(t=1)$ are smaller than $\phi_c$. As the energy is increased
to $E=0.0639$ some regular motions lose stability and chaotic
motions appear. They can be seen in the lower left corner of the
configuration space $(R_1,R_2)$. Highlighted ridges spiraling
around the central resonance region are associated with hyperbolic
periodic trajectories. They separate the main resonant zone with
the frequency ratios
$\omega_1/\omega_{\alpha}=2,~\omega_2/\omega_{\alpha}=2$ from
other higher order resonant zones. We notice that the FLI fields
do not provide a precise determination of the locations of
hyperbolic periodic orbits because its resolution is not sharp
enough and some regions do overlap. For the energies $E=0.07$ and
$E=0.08$ more chaotic trajectories appear and the overall dynamics
is more chaotic. This is indicated by the increasing number of
points with higher values of the FLI (the value of FLI reaches
$22$ for the $E=0.08$ ) in the configuration space.
We notice that at $E=0.07$ a bifurcation of the main resonance $2:2:1$ has occurred.
The central island which leads the regular dynamics for $E<0.07$ is split
into two parts. These regular parts are connected and survive for larger energies,
even near the dissociation threshold. They are associated with a resonance $1:4:1$.\\
For the energy $E=0.098$, only a few regular regions around
elliptic periodic orbits remain. They are separated by a large set
of initial conditions corresponding to chaotic motions. We
identify all the surviving resonances by computing the frequency
ratios for the trajectories generated inside the islands.

For energies close to  dissociation (see for instance
Fig.~$1(f)$), most of the Lyapunov field consists of initial
conditions associated with strongly chaotic motions~: The values
of the FLI $\phi(t=1)$ are larger than 16 except several resonant
islands and the percentage of regular behavior is below 30\%.
These remaining islands correspond to elliptic periodic orbits.
Large dark zones appear on th right hand side of the FLI fields
for $R_1> 4.8$. In order to get insight into the presence of low
values of the FLI for initial conditions in regions where we
expect a priori strongly chaotic motions, the time sequences for
$R_1$ and $R_2$ and the evolution of the FLI are plotted on
Figs.~\ref{fig:fig4a} and \ref{fig:fig4b} for two typical
trajectories with initial conditions such that $R_1>4.75$. Here
the FLI is defined as $\phi(t)=\log\Vert v(t)\Vert$ for
convenience. We observe that for the trajectory corresponding to
the dissociation of the molecule (see Fig.~\ref{fig:fig4a}) the
value of the FLI grows slowly (logarithmically). This is due to
the fact that C and O atoms are moving under the influence of
Morse potential $V_2$ mostly. Contributions from the other terms
in the potential are too small to affect the dynamics. The system
is very close to integrability. For the trajectory near the
dissociation (see Fig.~\ref{fig:fig4b}) we observe an intermittent
behavior in the time sequence for the $R_2$. Regular oscillations
are interrupted by higher amplitude peaks. Regular motion
corresponds to the intervals of time when the S atom is far from
the CO bond to affect the dynamics. During this interval of time
the quantity of the FLI is decreasing. This has a significant
effect on the resulting largest value of FLI at the end of the
interval of time. From the FLI fields, we observe that the region
with high values of the FLI is sharply joint at $R_1\in [4.7,
4.8]$ with the region where trajectories dissociate or demonstrate
an intermittent behaviour discussed above. At this stage, it is
not clear whether there is a surface barrier dividing phase space
into twoparts with completely distinct dynamics and what the
conditions are on an initial configuration of the OCS for the
trajectories to show one or the other type of dynamical behaviour.

We use the FLI field data to compute the percentage of regular versus irregular orbits for the system at different energies. We compare our results with an existing classification in Ref.~\cite{cart82}. The percentage of regular motions for different values of the energy $E$ is
plotted versus $E$ on Fig.~\ref{fig:perE}. Our computations give the bounds
for the critical energy for the onset of irregularities in the
system which are $0.05<E_c<0.0639$. This result
agrees with the value for the critical energy $E=0.0639$
calculated in Ref.~\cite{cart82}. However at the energy close to dissociation
($E=0.098$), we observe many more regular trajectories than the estimate given in Ref.~\cite{cart82}.
This can be due to the fact that in our studies we calculated FLI
for $132\times 110$ trajectories
whereas  the previous stability analysis was performed for 20
microcanonically distributed trajectories in Ref.~\cite{cart82}, and also to the fact that we consider initial trajectories in the configuration space (i.e.\ with zero kinetic energy). For the interval of energy $0.07<E<0.08$ we see a significant change in the dynamical behavior of the system. About 15\% of regular trajectories lose their stability. Therefore most of the important bifurcations in this system happen in this interval of energy.

\section*{Conclusions}
We have applied the Fast Lyapunov Indicator method in the
phase space of highly excited  planar OCS. This method is based on
an analysis of the linear stability of  trajectories. It gives
pictures of the dynamical regimes in  phase space, analogous to
Poincar\'e sections for systems with two degrees of freedom. In
conjunction with a search of periodic orbits, this method gives
insights into the dynamics of intramolecular energy flow in highly
excited molecules. We have applied this method to planar OCS with
energies below and at the dissociation threshold. The main results
are as follows: When the energy is well-distributed among the
three modes in the center of the configuration space, the behavior
is regular. When the bending mode is frozen and the two stretching
modes are in phase (O and S are vibrating in opposite phase)
chaotic behavior is seen. We have provided pictures of phase space
for different values of the energy below and at the dissociation
threshold~: It allows one to identify the mechanism of transition
to chaos and dissociation.

\newpage
\

{\bf FIGURES CAPTIONS}\\

{\rm Figure} $1$. \\
  FLI versus time curves for the OCS
system for the initial conditions for chaotic, weakly chaotic,
quasiperiodic and periodic orbits respectively:
     {a)} ( $2.762, 1.911$ ),\\
     {b)} ( $3.67, 2.307595$ ),
     {c)} ( $3.65, 2.307595$ ),
     {d)} ( $3.615178, 2.307595$ ).

\begin{figure}
\includegraphics*[width=14cm, height=14cm]{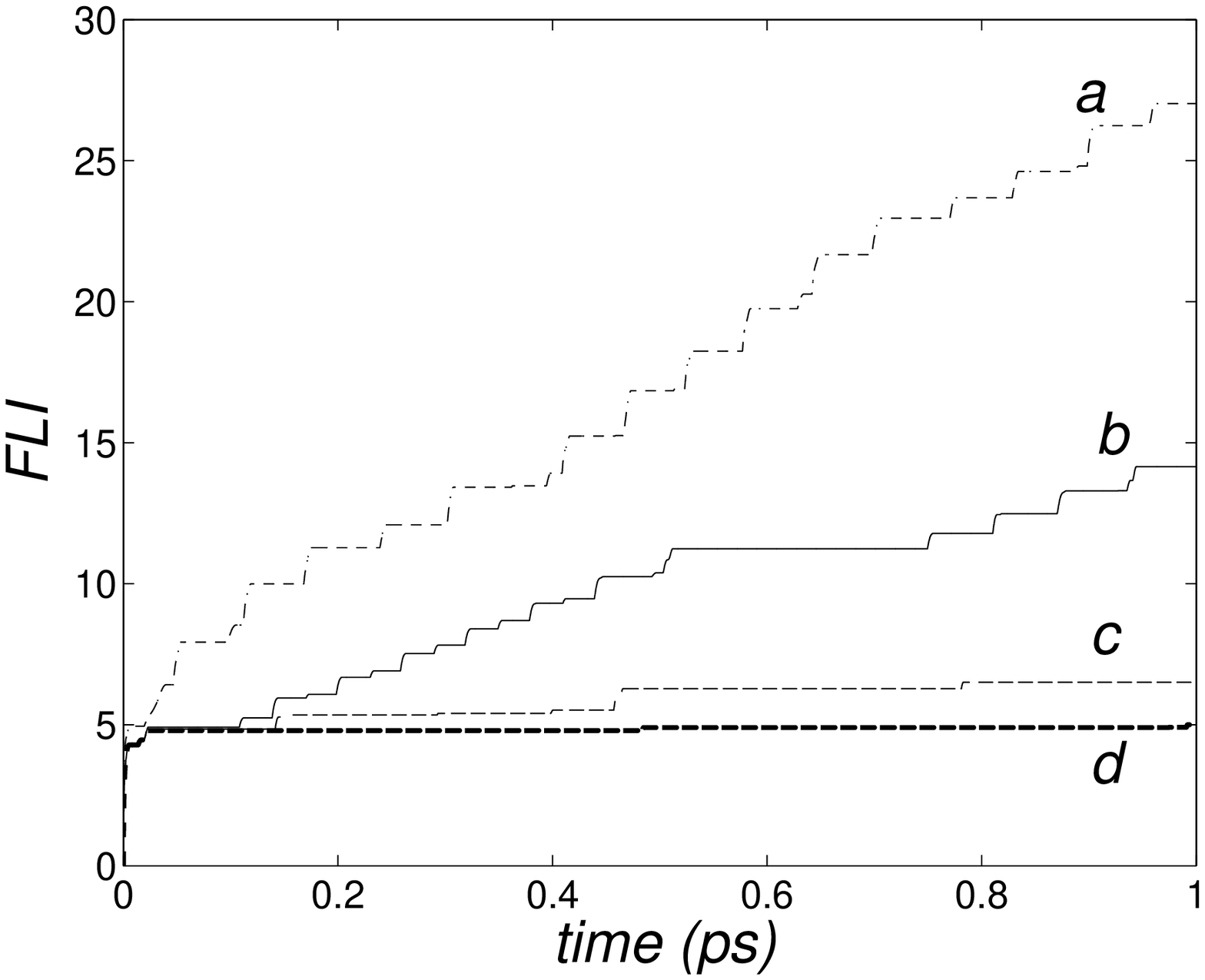}
\caption{}
\label{fig:FLIt1}
\end{figure}

\begin{figure}
  \includegraphics*[width=14cm, height=14cm]{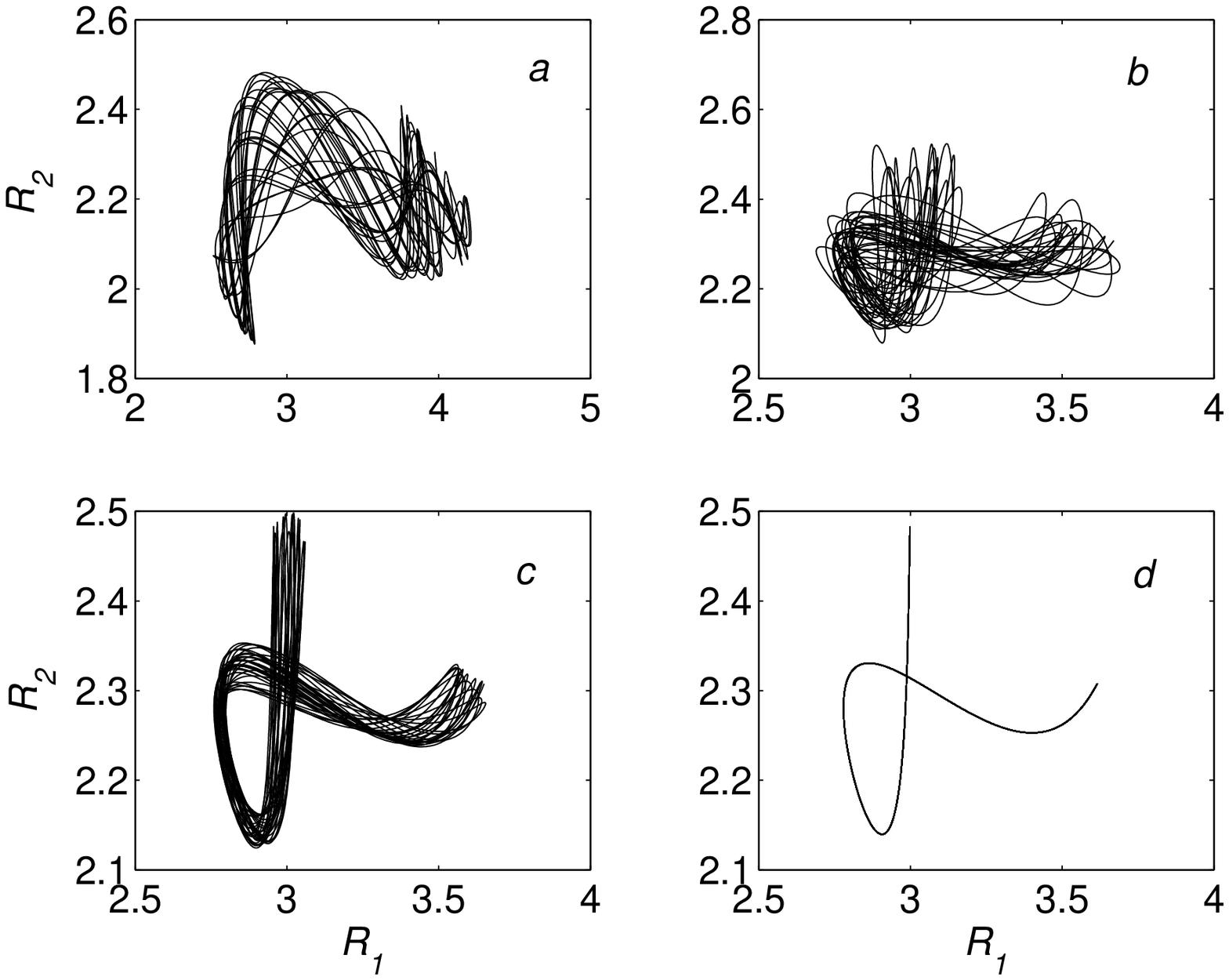}
\caption{}
\label{fig:FLIt1cs}
\end{figure}

\begin{figure}
  \includegraphics*[height=14cm,width=14cm]{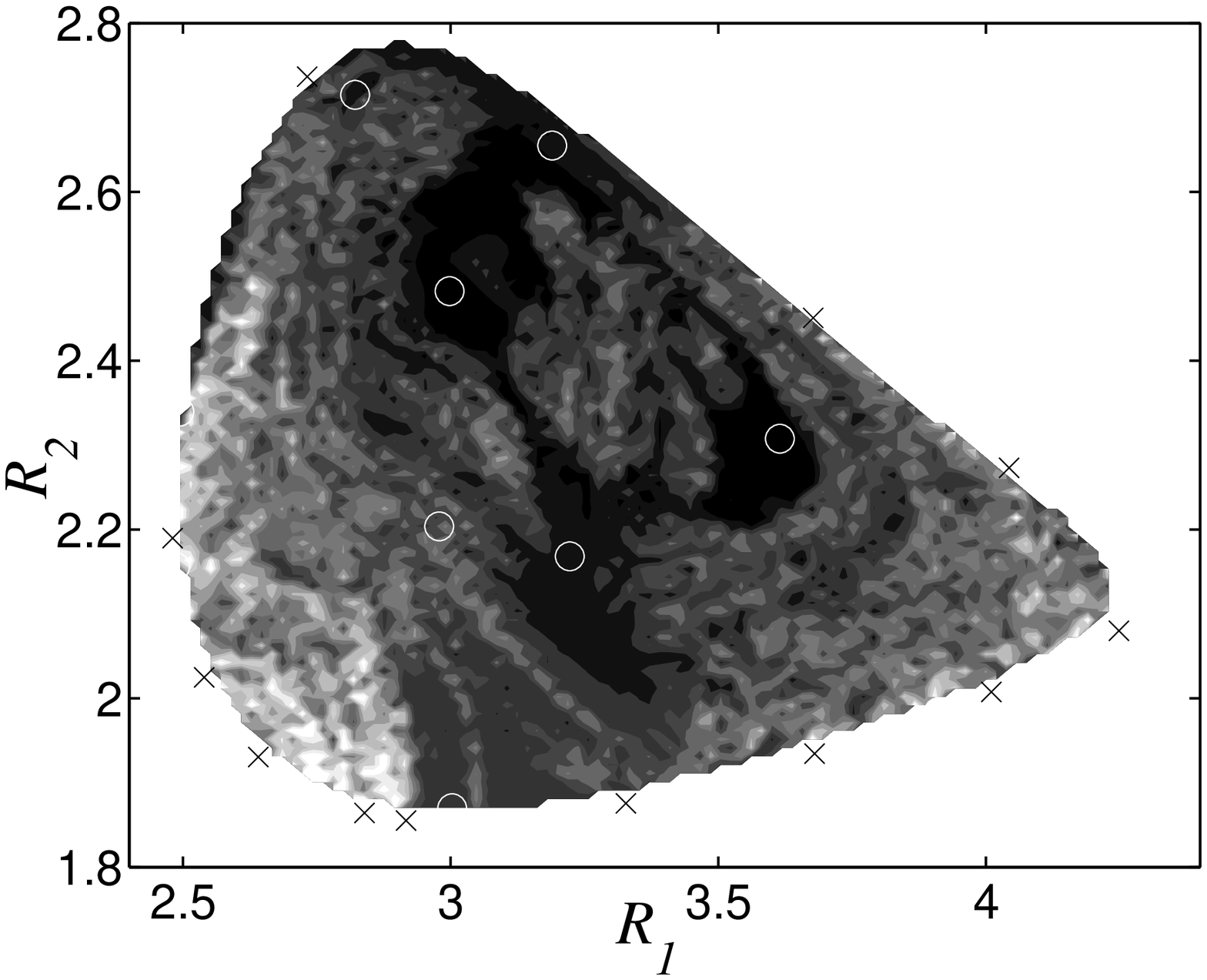}
\caption{}
\label{fig:FLIf1}
\end{figure}

\begin{figure}
  \includegraphics*[width=17cm, height=20cm]{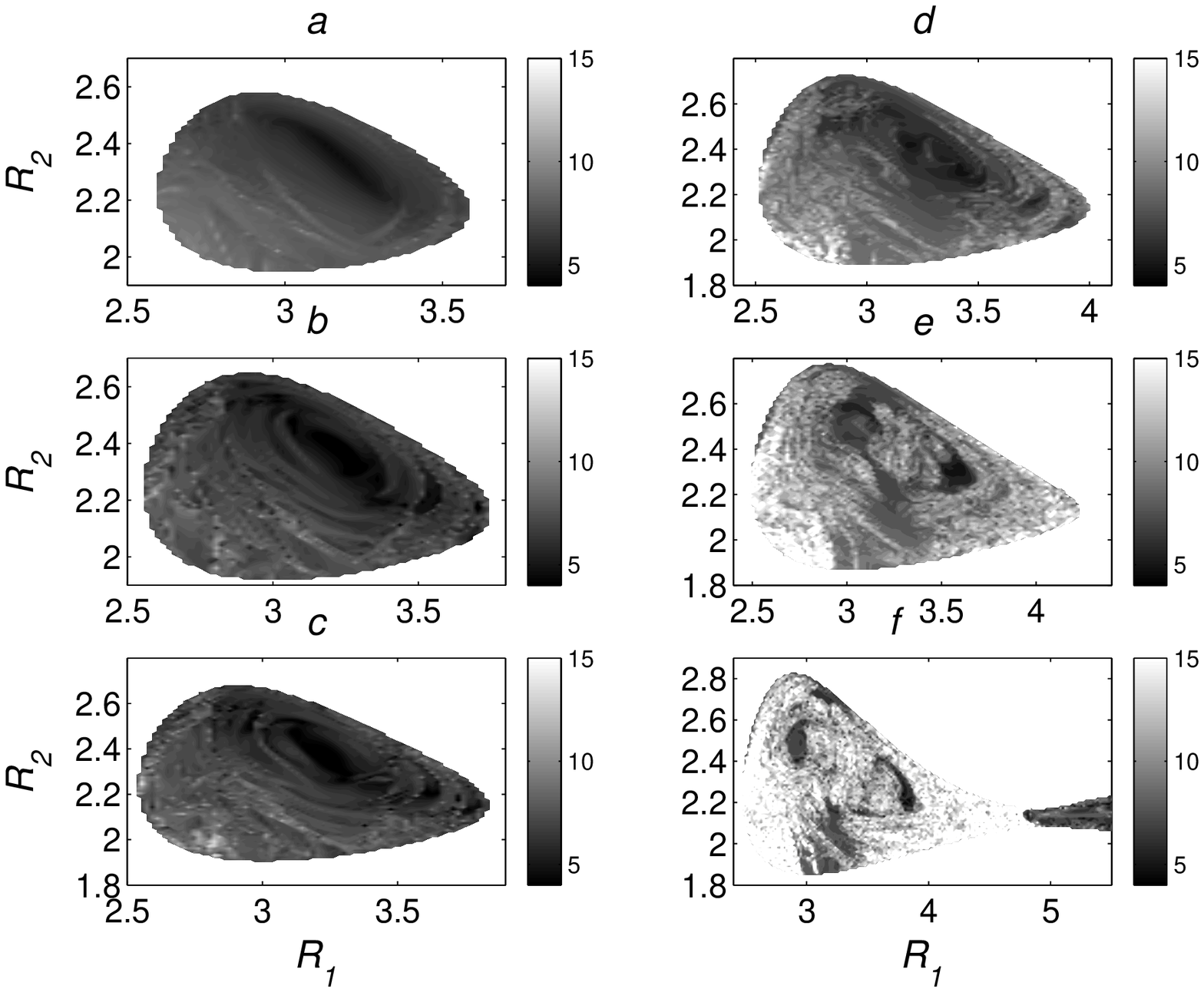}
\caption{}
\label{fig:FLIf2}
\end{figure}

\begin{figure}
  \includegraphics*[width=14cm, height=14cm]{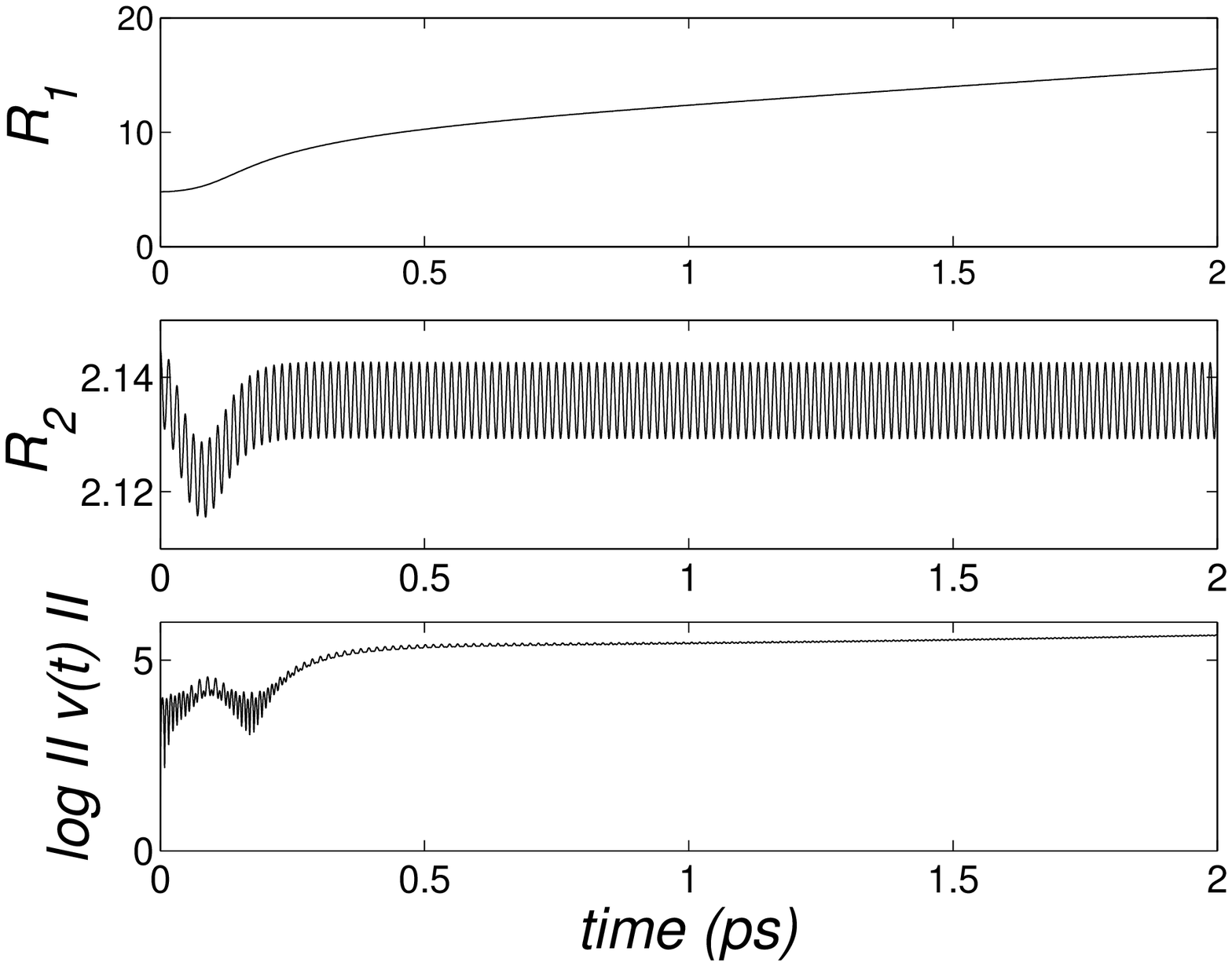}
\caption{}
\label{fig:fig4a}
\end{figure}

\begin{figure}
  \includegraphics*[width=14cm, height=14cm]{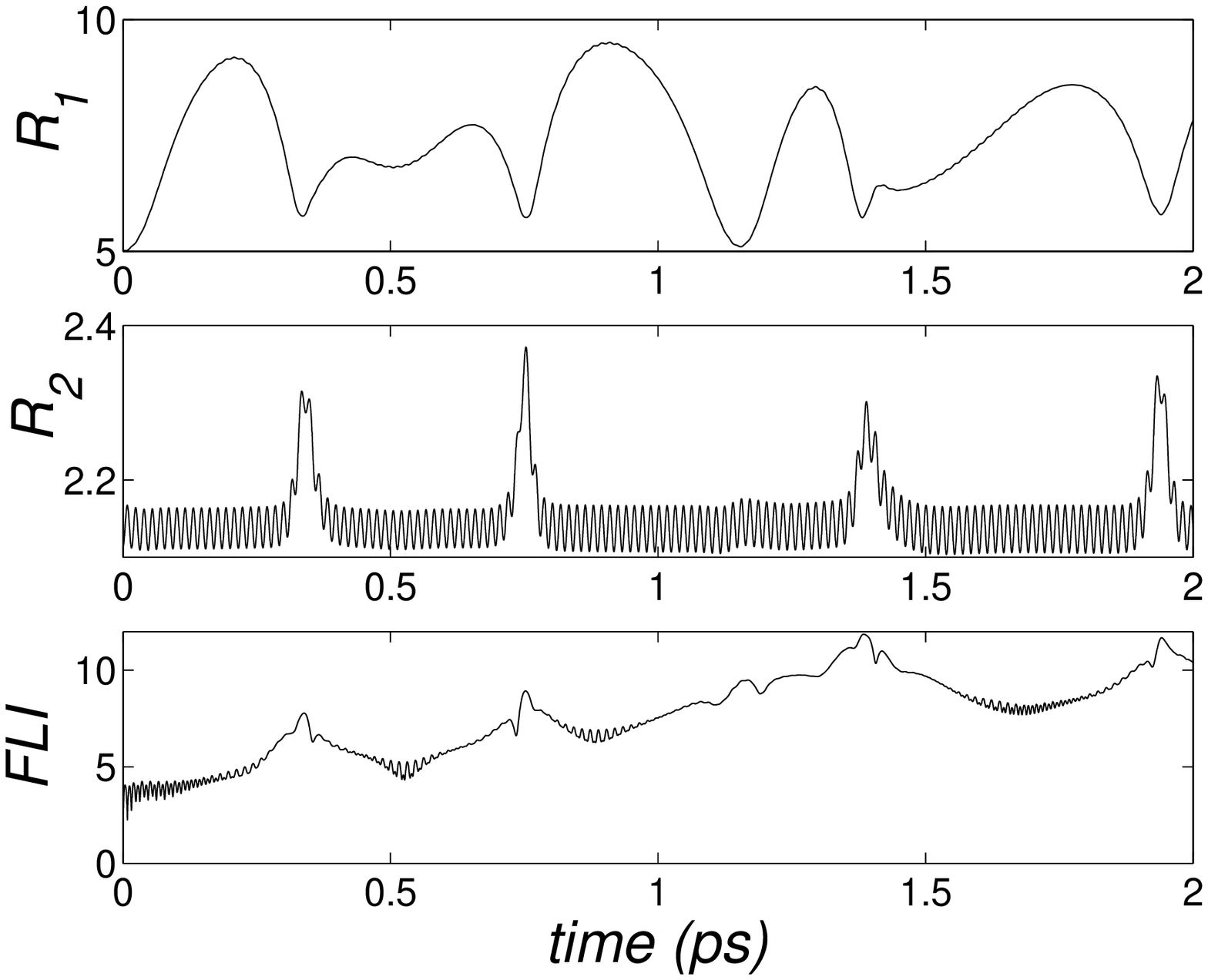}
\caption{}
\label{fig:fig4b}
\end{figure}

\begin{figure}
  \includegraphics*[width=14cm, height=14cm]{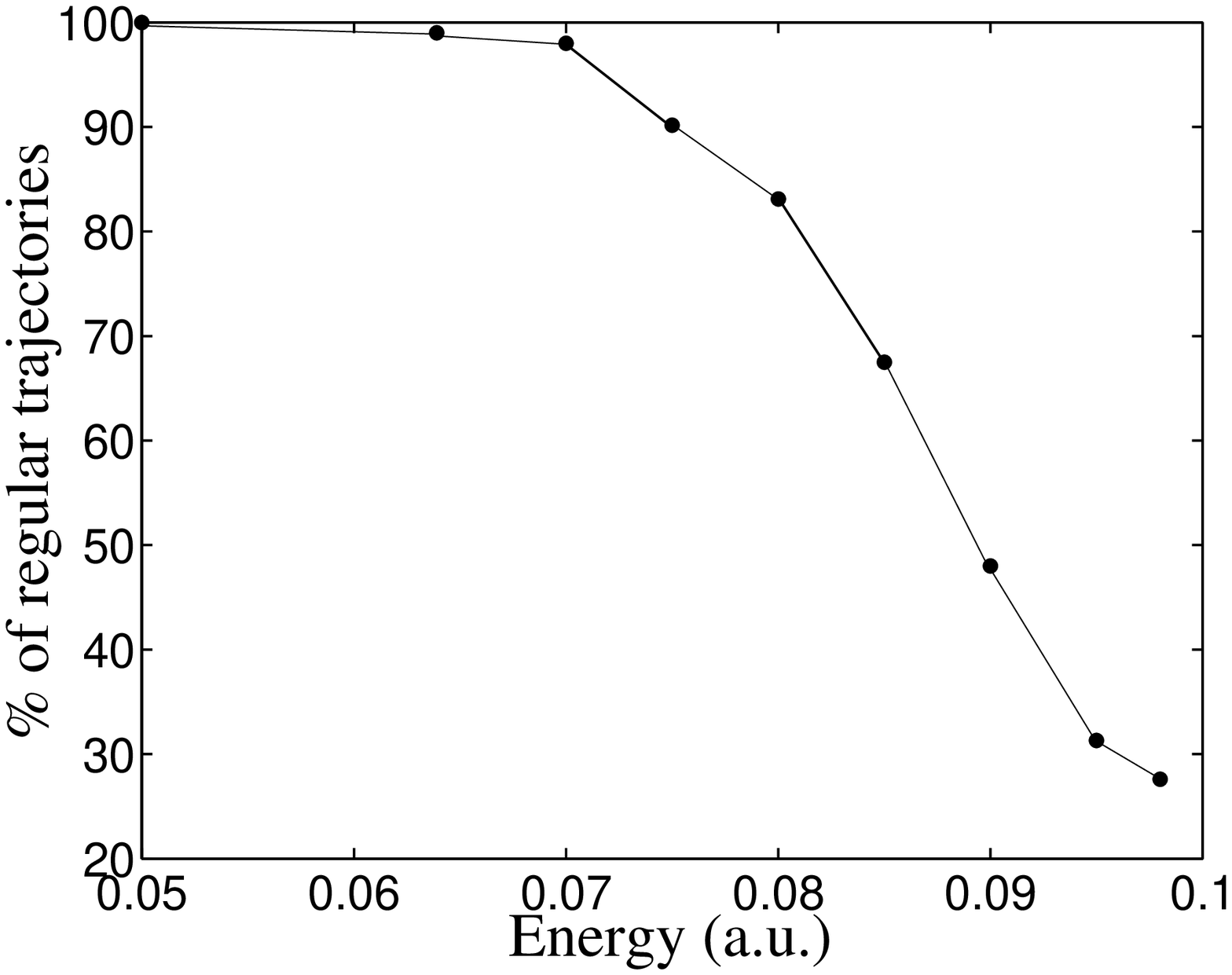}
\caption{}
\label{fig:perE}
\end{figure}

{\rm Figure} $2$. \\
Trajectories in the configuration space for the three initial
conditions of Fig.~\ref{fig:FLIt1}. Trajectories $(b)$ and $(c)$
are chosen in the vicinity of the periodic orbit $O_a$ which is
represented by the trajectory $(d)$. The time of integration is
1ps.

{\rm Figure} $3$.\\
Contour plot for the FLI values in the configuration space
$(R_1,R_2)$ of the OCS molecule for $E=0.09$. The circles
represent elliptic periodic orbits and crosses represent collinear
hyperbolic periodic orbits.

{\rm Figure} $4$\\
Contour plots for the FLI values in the configuration space
$(R_1,R_2)$ of the OCS molecule given at the energies:
               {a)} $E=0.05$,
               {b)} $E=0.0639$,
               {c)} $E=0.07$,
               {d)} $E=0.08$,
               {e)} $E=0.098$,
               {f)} $E=0.1$.

{\rm Figure} $5$ \\
Time series of $R_1$, $R_2$ and the FLI evolution for the
trajectory at the dissociation energy \\ $E=0.1$. Initial
condition : $R_1 = 4.8$, $R_2 = 2.145$.

{\rm Figure} $6$\\
Time series of $R_1$, $R_2$ and the FLI evolution for the
trajectory near the dissociation energy. Initial condition : $R_1
= 5.02$, $R_2 = 2.11$.

{\rm Figure} $7$\\
Percent of regular trajectories versus energy for the OCS.
The interval of energy is \\ E $\in$ [  $0.05, 0.098$  ].\\

\begin{table}
                \begin{tabular}[t]{|c|c|c|c|c|}
                \hline
                Name & Projection & $T$ (ps) & $(R_1,R_2)$ & $m:n:k$\\
                & $(R_1,P_1)$ and $(R_2,P_2)$ & & & \\
                \hline
                $O_a$ & \includegraphics*[width=6cm, height=3cm]{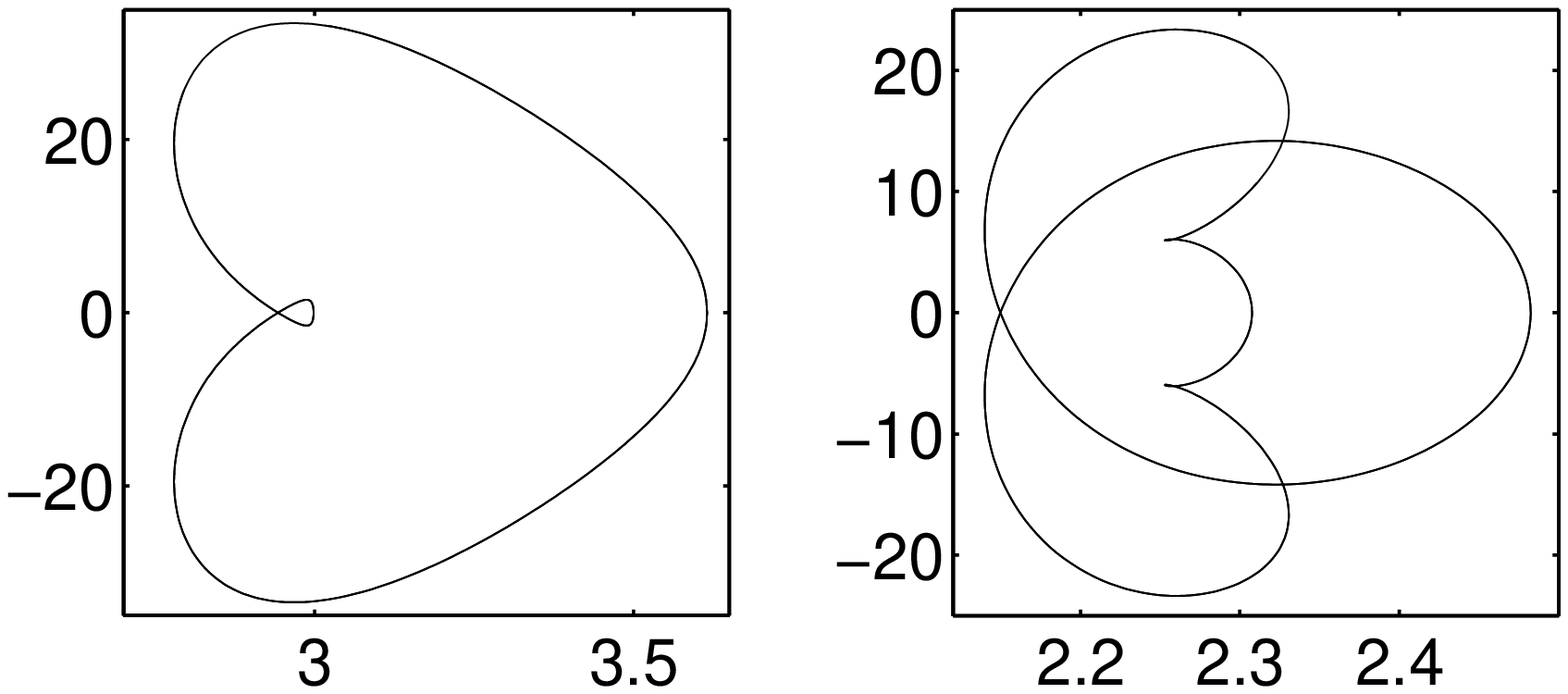} & 0.063 & $(3.615178,~2.307595)$ & $1:1:1$ \\
                & & & $(2.998235,~ 2.482408)$ & \\
                \hline
                $O_b$ & \includegraphics*[width=6cm, height=3cm]{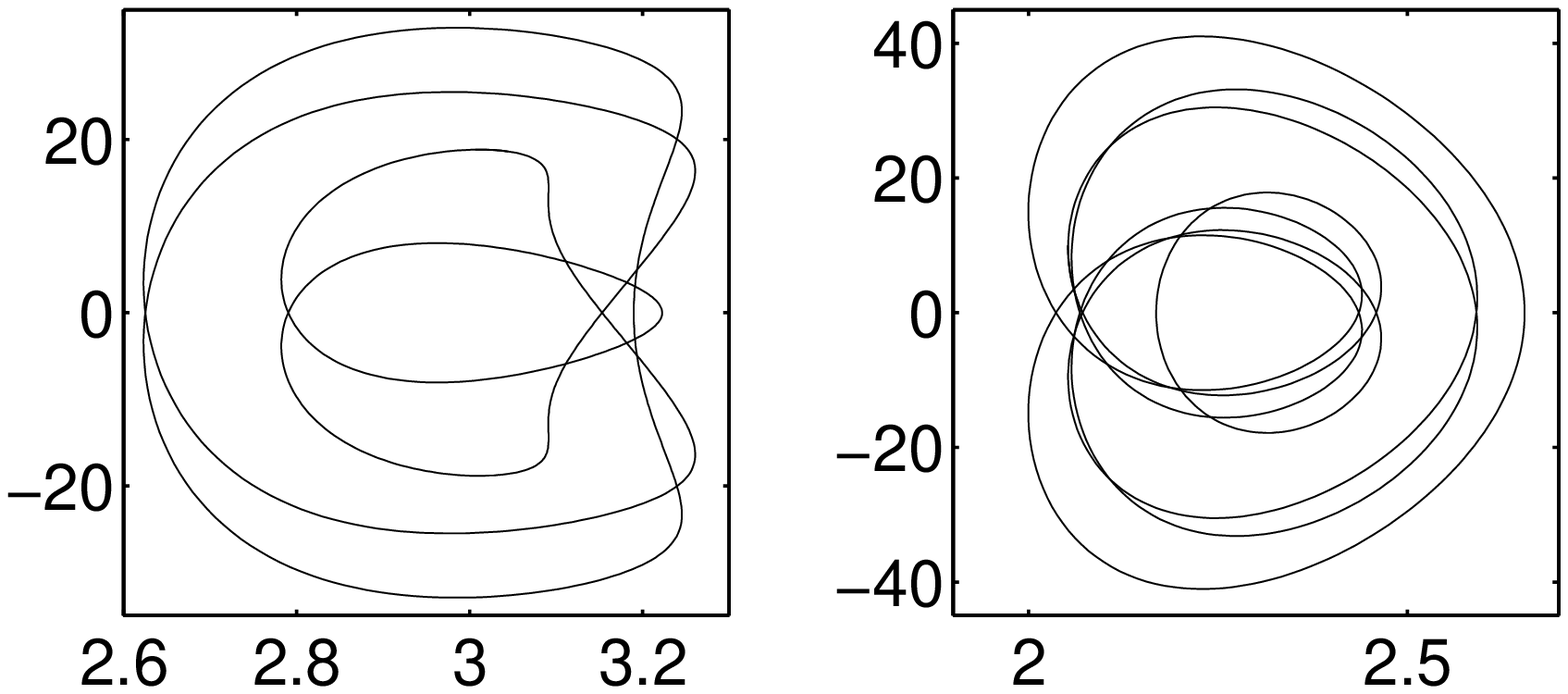} & 0.122 & $(3.189711,~ 2.654841)$ & $2:3:1$ \\
                & & & $(3.222491,~ 2.168308)$ & \\
                \hline
                $O_c$ & \includegraphics*[width=6cm, height=3cm]{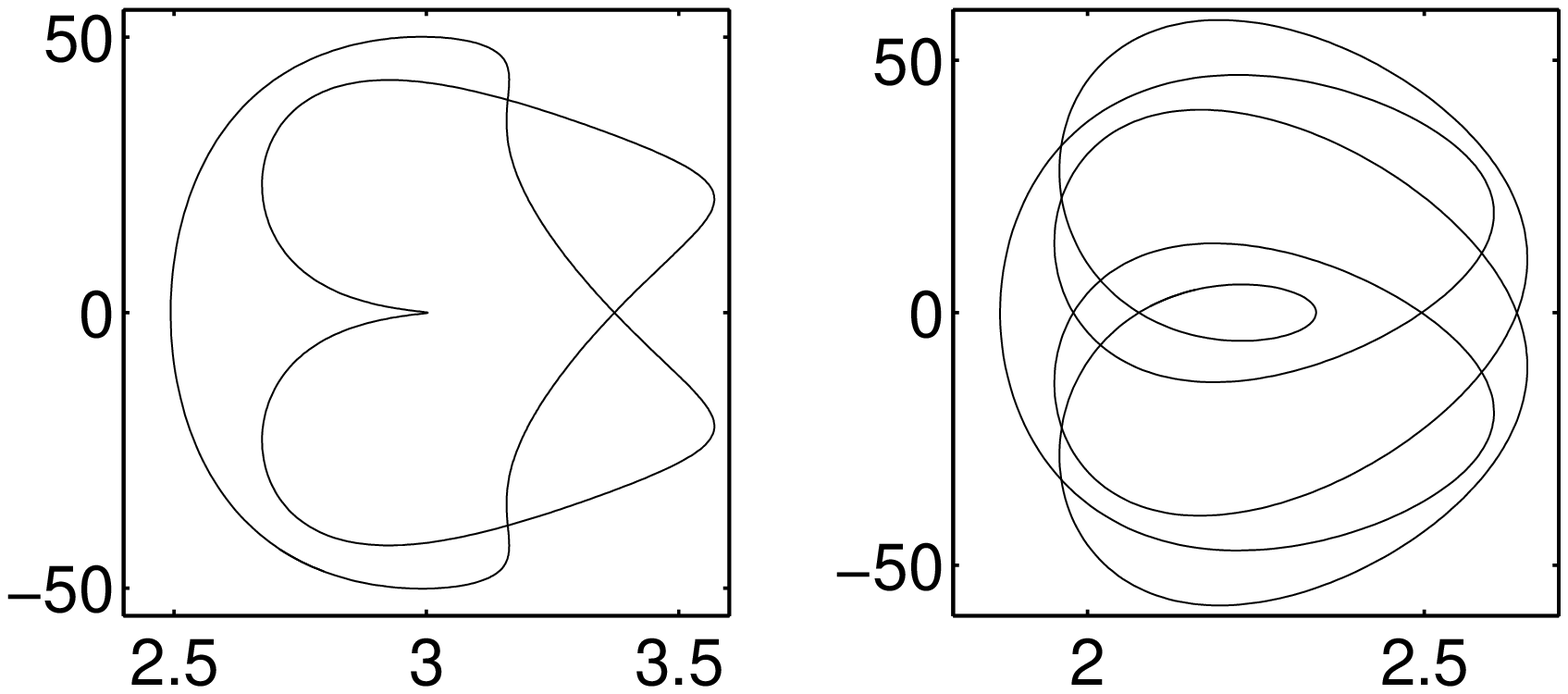} & 0.087 & $(2.493375,~ 2.339454)$ & $2:2:1$ \\
                & & & $(3.002806,~ 1.869968)$ & \\
                \hline
                $O_d$ & \includegraphics*[width=6cm, height=3cm]{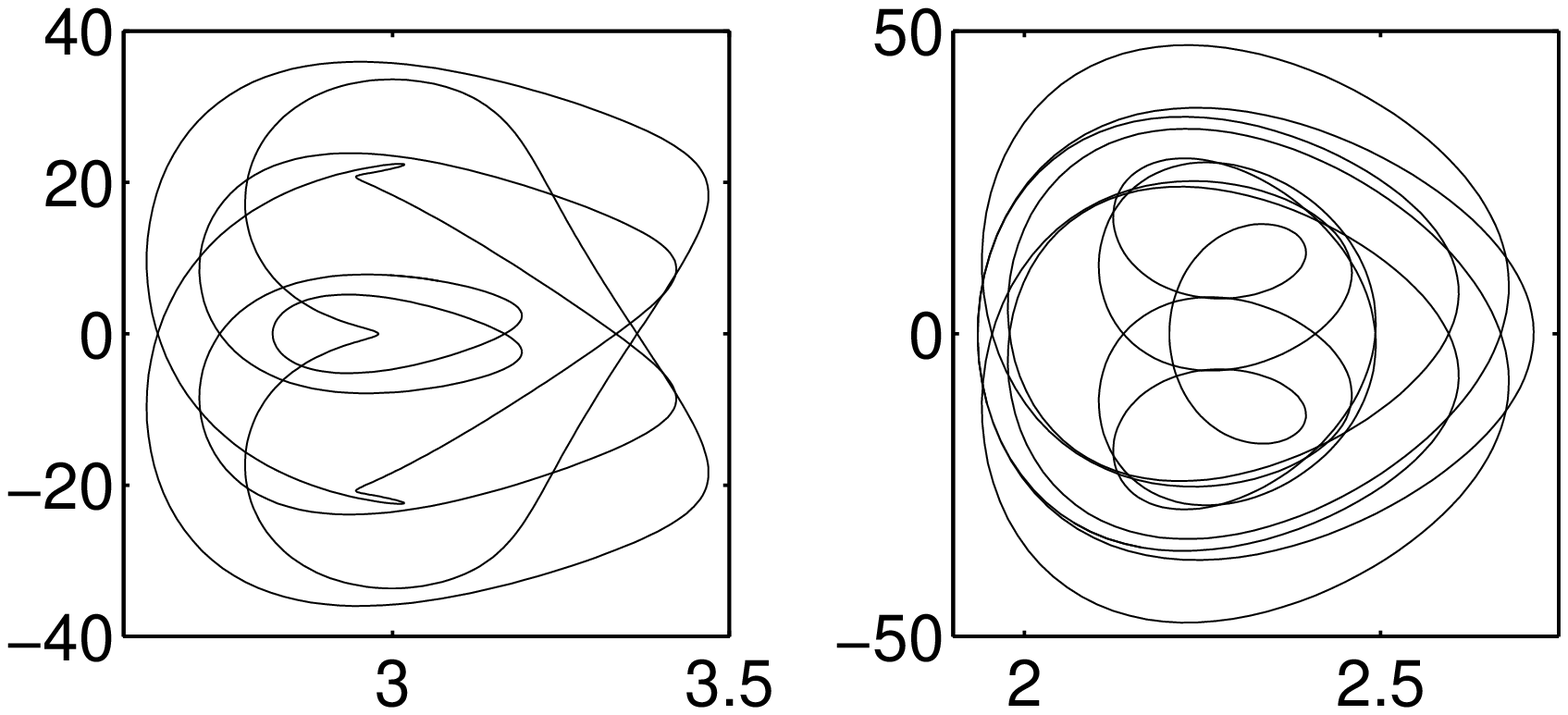} & 0.195 & $(2.821580,~ 2.714920)$ & $4:11:3$ \\
                & & & $(2.978729,~ 2.203665)$ & \\
                \hline
                $O_e$ & \includegraphics*[width=6cm, height=3cm]{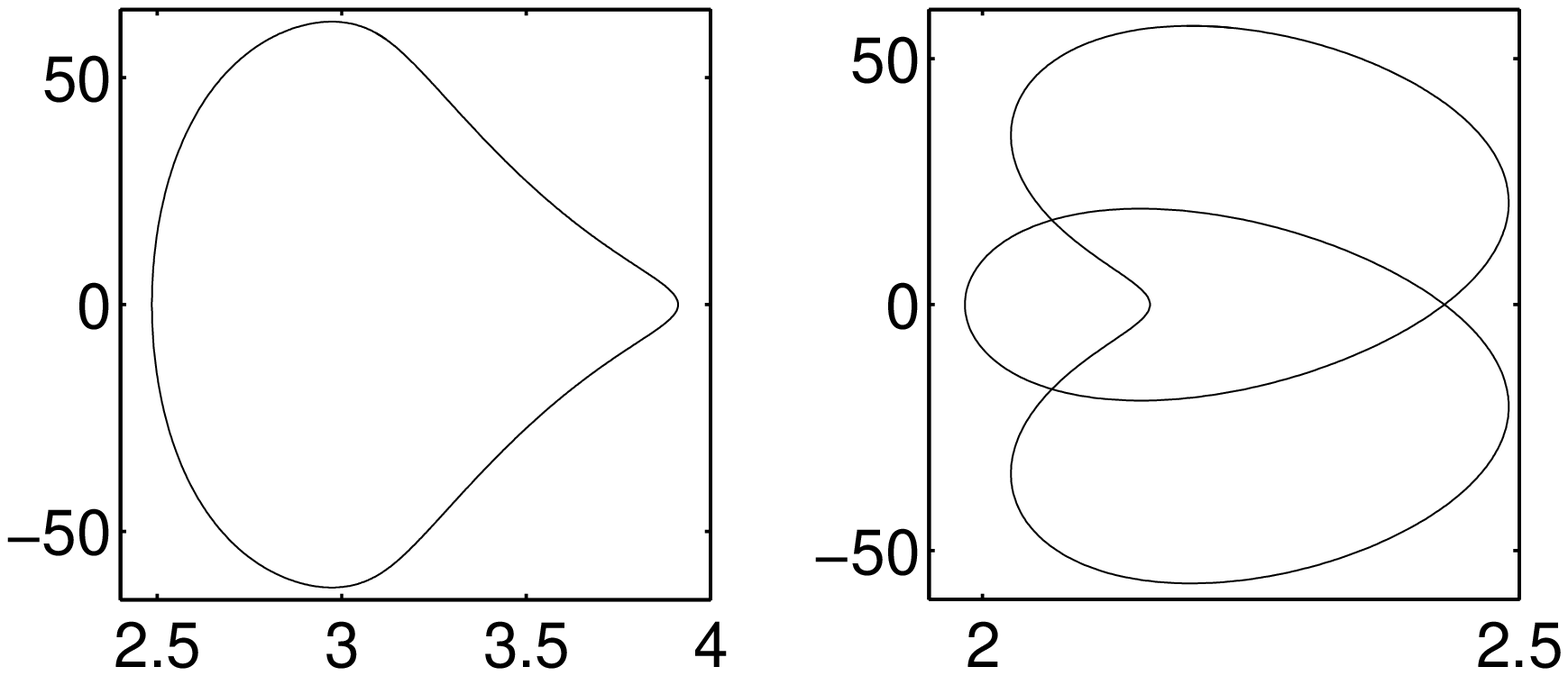} & 0.049 & $(3.911413,~ 1.983623)$ & $1:1:0$ \\
                & & & $(2.485514,~ 2.155984)$ & \\
                \hline

                \end{tabular}

        \caption{Characteristics of the main elliptic periodic orbits~: two projections on the planes $(R_1,P_1)$ and $(R_2,P_2)$ are given. We also provide the period $T$ expressed in picoseconds and the two points which are the intersections of the orbits with the configuration space. }
        \label{tab:op}
\end{table}

\newpage

\newpage
\newpage
\newpage
\newpage
\newpage
\newpage

\end{document}